\documentclass[10pt]{article}
\usepackage{amsmath}
\usepackage{amsthm}
\usepackage{amsfonts}
\usepackage{amssymb}
\usepackage{dsfont}
\usepackage{mathrsfs}
\usepackage{graphicx}
\usepackage{epstopdf}
\usepackage[body={17.5cm, 22cm},right=2.2cm]{geometry}
\newcommand{\ie}{{\emph{i.e.}}~}
\usepackage{braket}
\newcommand{\oket}[2]{|#1 )_{#2}}
\newcommand{\obra}[2]{ \phantom{}_{#2}(#1 |}

\newcommand{\SU}{\mathrm{SU}(2)}
\newcommand{\SL}{\mathrm{SL}(2,\mathbb{C})}
\newcommand{\lsq}{L^2(\SU)}
\newcommand{\su}{{\mathfrak{su}}(2)}
\newcommand{\DD}{\mathcal{D}}
\newcommand{\EE}{\mathcal{E}}
\newcommand{\XX}{\mathscr{D}}
\newcommand{\KK}[2]{\mathscr{K}^{#1}_{(#2)}}
\newcommand{\HK}[2]{\mathrm{K}^{#1}_{(#2)}}
\newcommand{\FT}{{\mathscr{F}}}
\newcommand{\stp}[1]{\star_{#1}}

\newcommand{\Tr}{\mathrm{Tr}}

\newcommand{\iu}{\mathrm{i}}

\begin{document}

\setcounter{page}{0}
\thispagestyle{empty}
\begin{center}

~

\vspace{1.cm}

{\Large
\textbf{New coherent states and modified heat equations
}}\\[1cm]

{\large  Antonio Pittelli$^{a}$}
and
{\large  Lorenzo Sindoni$^{b}$}
\\[0.5cm]

{\small \textit{$^{\rm a}$ Universit\`a degli Studi di Trieste\\
Via A.Valerio  2, 34127 Trieste, Italy\\}}
{\small \textit{$^{\rm b}$ Max Planck Institute for Gravitational Physics,\\
Albert Einstein Institute,\\
Am M\"uhlenberg 1, 14476 Golm, Potsdam, Germany}}

\end{center}

\vspace{.8cm}
\vspace{0.3cm} {\small \noindent \textbf{Abstract} \\[0.3cm]
\noindent
 We clarify the relations between certain new coherent states for loop quantum gravity and the analytically continued
heat kernel coherent states, highlighting the underlying general construction, the presence of 
a modified heat equation as well as the way in which the properties of the heat kernels are automatically
inherited by these new states. }


\section{Introduction}

In \cite{OPS1}, extending previous work on semiclassical states in Loop Quantum Gravity (LQG) \cite{Thomas1,Thomas2,Thomas3,Thomas4,ThomasComplexifier}, a new family of coherent states 
(for the group $SU(2)$) has been introduced.
Both choices present some advantages: for instance, $\SL$ heat kernels have got very simple convolution properties; on the other hand, flux coherent states provide correction-free expectation values for the flux operator $\hat E$. Therefore these states might
provide an alternative, possibly more transparent, semiclassical treatment of the quantum operators associated to
the classical phase space on which Loop Quantum Gravity \cite{RovelliBook} is based.

These new states are also naturally formulated in the flux representation of LQG, introduced 
in \cite{fluxrep}, where, using the concept of non-commutative Fourier transform \cite{FreidelMajid,JoungMouradNoui},
a representation of the LQG building blocks based on functions over (several copies of) the $\su$ Lie algebra flux variables is developed.

The purpose of this paper is very specific: we want to reconsider the results of \cite{OPS1}, with emphasis on the 
relationship with the coherent states they were originated from \cite{Thomas1,ThomasComplexifier} using the techniques
of complexifiers and the analytic continuation of heat kernel states on $\SU$.
We will present the results in the group representation, so that the connection is even more clear.
We will show, in particular that the states introduced in \cite{OPS1} are indeed heat kernels on their own
right, associated to a different choice of heat equation, in particular a different Laplace operator.
Not only this observation makes more transparent the results discussed in \cite{OPS1}, but also shows
how to construct further generalizations.

In this paper we will restrict the attention to the classical phase space of a single holonomy/flux pair, 
whether this is associated to a single link of a graph or to a more extended structure \cite{STW,OPS2}.
The holonomy flux algebra is given by
\begin{equation} 
\left[ \, \hat{E}^i,\hat{h}\right] = i \hbar (8\pi G\gamma)  R^i \triangleright \hat{h}, \qquad \left[ \, \hat{E}^i,\hat{E}^j \, \right] = i  \hbar (8\pi G\gamma)  \epsilon^{ij}_k \hat{E}^{k},   
 \end{equation}
where $\hbar (8\pi G\gamma) = 8\pi l_p^2 \gamma$ has the dimension of a length squared\footnote{We are using units in which $c=1$.} and $\hat{R}^{i}$ is the i-th 
right invariant vector field on $\SU$.
For the present discussion we will not be concerned with dimensionful quantities, and hence
we will use rescaled flux operators such that all the relevant variables are dimensionless.

Wavefunctions for kinematical states can be constructed by suitable choices of representations of several copies of this algebra (and by imposition of all the gauge invariances associated to
gravity), thus leading to the concept of cylindrical functions \cite{RovelliBook} (see also \cite{Carlos} for a discussion of the feasibility 
of the construction within the flux representation). We will not embark in the construction of general
states. Rather, we will focus on a single copy of this algebra of operators.

We will then construct wavefunctions as square integrable functions on one copy of the group, or,
alternatively, on the Lie algebra $\su$. We will work in the group representation, with the following representation for the flux operators:
\begin{equation}
E^{j}(f(h)) = \lim_{\epsilon\rightarrow 0} \left[ i \frac{d}{d\epsilon} f\left( e^{-i\sigma^j\epsilon} h \right) \right].
\end{equation}

Heat kernels are the most natural generalization of the Gaussian functions to the case of more general (Riemannian) spaces. They are defined in terms of the heat equation
\begin{equation}
\partial_t \rho^{t}(x;y) = \triangle_{x} \rho^{t}(x;y); \qquad \rho^{t=0}(x;y) = (-g(x))^{-1/2} \delta(x-y),
\end{equation}
where the Laplacian operator $\triangle$ depends on the metric tensor, and $t$ is an evolution
parameter.
Obviously, $\SU$ can be seen as a Riemannian space (indeed, as $S^{3}$ with the canonical
metric on it). Therefore, it is possible to naturally define a heat kernel on the group \cite{Camporesi}, as solutions
of the heat equation
\begin{equation}
\partial_t K^{t}(gg_{0}^{-1}) = \triangle_{\SU} K^{t}(gg_{0}^{-1}); \qquad K^{t=0}(gg_0^{-1}) = 
\delta(gg_{0}^{-1}) \label{eq:heatk},
\end{equation}
where we are using the Laplace--Beltrami operator on $\SU$,
\begin{equation}
\triangle_{\SU} =- \delta_{ij} E^{i}E^{j}.
\end{equation}

The Dirac delta on the group, providing the initial condition for the differential equation, completes the definition of these functions. 
The heat kernel $K^{t}$ can be used to construct coherent
states, with the appropriate identification of $t$ with the spread of the wavefunction and of
$g_0$ with the peak, its maximum. By construction,
\begin{equation}
\langle E^{i} \rangle = 0,
\end{equation}
on these states, and therefore we need an additional step in order
 to define a state that is coherently peaking on a generic
phase space point, labeled by $g_{0} \in \SU$ and by $x_0\in \su$.

In order to use them to construct semiclassical states peaking on a given point of the single link classical
phase space, in
\cite{Thomas1,Thomas2,Thomas3,Thomas4,ThomasComplexifier} these functions
(seen as functions of the peak element $g_0$) were analytically continued, turning 
the heat kernel \eqref{eq:heatk} into a function on $\SL$,
\begin{equation}
\HK{t}{g_0}(h)= K^{t}(hg_{0}^{-1}), \qquad g_{0} \in \SU \rightarrow \HK{t}{G_0}(h), \qquad G_0 \in \SL.
\end{equation}
We refer to the literature for a detailed discussion of the properties of the states defined in
such a way. See also \cite{HallMitchell,HallMitchellErratum,HallMitchell2} for further discussions
on related results.

The approach discussed so far gives particular relevance to the natural isomorphism between the phase space constructed on a single
copy of $\SU$, $T^*\SU$, and the analytic continuation of $\SU$ itself, namely $\SL$.
In \cite{OPS1} a different path has been followed, exploiting the point of view given by the definition
of a Fourier transform on the group \cite{FreidelMajid,JoungMouradNoui} (see also \cite{Matti}
for a discussion of possible extensions to more general Lie groups).
We will denote this operation as $\FT:\lsq \rightarrow L^2(\su,\star)$, mapping square integrable functions on the group 
onto square integrable functions on the Lie algebra $\su$, endowed with the ordinary Lebesgue measure but with a $\star-$product replacing the ordinary commutative product of functions.

As shown in \cite{OPS1}, by modifying heat kernels with the multiplication by a plane wave,
\begin{equation}
\KK{t}{g_0,x_0}(g) = e_{g}(x_0)   \HK{t}{g_0}(g),
\end{equation}
one obtains states that have essentially the properties of coherent states, with respect
to averages, peakedness properties, completeness and overlap. Here, plane waves are defined
as complex valued functions over $\SU\times \su$:
\begin{equation}
(g,x) \in \SU \times \su \mapsto e_{g}(x) \stackrel{\mathrm{def}}{=} \exp\left( - \frac{i}{2}
\mathrm{Tr}\left( |g| x \right),
\right),\qquad |g|=\mathrm{sign}(\chi(g))g,
\end{equation}
where the group element $g$ is assumed to be represented by its fundamental representation matrix acting on $\su$, in order for the expression above to make sense.

The results of \cite{OPS1} are based on the fact that a) the states originate from a heat kernel and b) that
in terms of the Fourier transform, this state is just the ordinary heat kernel translated by $x_0$
in the Lie algebra representation, \ie
\begin{equation}
\FT\left(\KK{t}{g_0,x_0}\right)(x) = \FT\left(\HK{t}{g_0}\right) (x-x_0).
\label{eq:hkfourier}
\end{equation}

The results were established by means of an explicit analysis of the various integrals which
are involved in the calculations of the various expectation values and overlap functions. However, it was
not clear whether it is possible to extend the results of the complexifier construction and the
proof of the existence of an annihilation operator to this particular case, even though equation
\eqref{eq:hkfourier} suggests that this might be the case. 
In the next sections we will give a detailed derivation of these states $\KK{t}{g_0,x_0}$ as heat kernels
of suitably defined Laplace operators, thus making manifest the correspondence with the structure of the 
standard heat kernel coherent states, a cleaner and more automatic derivation of the results of
\cite{OPS1} and a recipe for further generalizations.


\section{Remarks on translations}

As we have seen, one of the key ideas is to use the multiplication with a plane wave in order
to achieve a translation in momentum space. Here we will give a more abstract description of this operation, which will
give us the opportunity to better grasp the structure of the reasoning, of the results and to offer a 
path for possible future
generalizations. Given $x_0\in \su$, define the linear map
\begin{equation}
U_{x_0} : \lsq \rightarrow \lsq; \qquad U_{x_0}(f)(h) := e_{h}(x_0) f(h). 
\end{equation}
This map is obviously unitary. Indeed, it is invertible, with inverse
$
U_{x_0}^{-1} = U_{-x_0},
$
since it
preserves the canonical scalar product of $\lsq$
\begin{equation}
\langle U_{x_0} f_{1} |U_{x_0} f_{2} \rangle = \langle f_{1} | f_{2} \rangle ,
\end{equation}
and its adjoint is the inverse:
\begin{equation}
U_{x_0}^{\dagger} = U_{-x_0} = U_{x_0}^{-1}.
\end{equation}

Therefore, if we have an orthonormal basis for $\lsq$, $f_{n}$ with $n$ in some index space,
the functions
\begin{equation}
f_{n}^{U} = U_{x_0} f_{n}
\end{equation}
will be an orthonormal basis for $\lsq$ too\footnote{The key of the argument is that, given the unitarity of $U$, Cauchy sequences are mapped onto Cauchy sequences, since the 
scalar product, and hence the norm and the distance derived from it are all conserved.}.
In particular, the Peter--Weyl theorem ensures us that the set of Wigner matrices is such
a basis. Hence, also
\begin{equation}
\DD^{j}_{ab}(h) = (U_{x_0}D^{j}_{ab})(h), \qquad j = 0,\frac{1}{2}, 1, \ldots
\end{equation}
are an orthonormal set of functions. Indeed, while orthonormality is automatically guaranteed by 
unitarity, their role as basis functions is also clear from elementary considerations.
For any $F \in \mathcal{L}^2(\SU)$, by construction, also $U_{x_0}^{-1}F$ belongs to
this space. Therefore,
\begin{equation}
\left(U_{x_0} F\right)(g) = \sum_{j} M^{j}_{ab} D^j_{ba}(g) =\sum_{j} M^{j}_{ab} 
U_{x_0}^{-1}
\DD^j_{ba}(g) = U_{x_0}^{-1} \sum_{j} M^{j}_{ab} 
\DD^j_{ba}(g),
\end{equation}
or
\begin{equation}
F(g) = \sum_{j} M^{j}_{ab} 
\DD^j_{ba}(g),
\end{equation}
for some matrices $M^{j}_{ab}$ (of course, the sum might be, in fact, a series).

Of course, this result can be seen in a easier way when everything is written in terms of noncommutative
functions on $\su$, the multiplication by the plane wave being just a translation.

It is clear that the construction of the unitary operator $U$ can be generalized to the case
in which the function that is used to define it is not necessarily a plane wave, but a sufficiently
regular phase factor. All the results enumerated so far will carry over to the general case,
with the exception of the picture in the algebra representation, where the translation
will be replaced by a convolution. Define the Fourier transform of a general phase factor:
\begin{equation}
\int_{\SU} dh \, \overline{e_{h}(x)} \exp(i\alpha(h)) = u_{\alpha}(x).
\end{equation}
Then, the Fourier transform of the product of a phase factor with a function on $\SU$ will be
\begin{equation}
\int_{\SU}dh \overline{e_h(x)}\exp(i\alpha(h)) f(h) = \int_{\SU} dh \int_{\su}d^3y\int_{\su}d^3z
\overline{e_h(x)}u_{\alpha}(y) e_{h}(y) e_{h}(z)  \FT(f)(z),
\end{equation}
whence we obtain the following convolution of Fourier transforms
\begin{equation}
\FT(Uf)(x) = \int_{\su}d^3 y\, u_{\alpha}(y) \stp{y} \FT(f)(x-y).
\end{equation}

We can easily conclude, then, that the flux coherent states introduced in $\cite{OPS1}$ are just the image of the heat kernel
coherent states\footnote{Here we are assuming that we are not analytically continuing them.} under a family of these maps :
\begin{equation}
\KK{t}{g_0,x_0} = U_{x_0}\HK{t}{g_0}.
\end{equation}
We will now use this to show what are the structures that are behind these new states, and to
what extent we can generalize the construction of coherent states following this path.


\section{The new coherent states as heat kernels}

In the group representation, the flux operator acts as a derivative operator; especially, it obeys the
standard Leibniz rule
\begin{equation}
E^{i}(f(h)g(h)) = (E^{i}f(h)) g(h) + f(h) (E^{i}g(h)) .
\end{equation}

In particular, we can compute the action of the fluxes on the plane waves: 
\begin{equation}
E^{j}(e_{g}(x_0))= 
\lim_{\epsilon\rightarrow 0} \left[ i \frac{d}{d\epsilon} \exp\left( -\frac{i}{2} \Tr\left\{ e^{-i\sigma^j\epsilon} |h| \sigma_k \right\} x_{0}^{k} \right) \right] = 
-\frac{i}{2}\Tr\left( \sigma^j |h| \sigma_k \right)x_{0}^{k}(e_{h}(x_0)) = v^{j}_{x_0}(h)e_{h}(x_0),
\end{equation}
where
\begin{equation}
v^{j}_{x_0}(h)=-\frac{\iu}{2}\Tr\left( \sigma^j |h| \sigma_k \right)x_{0}^{k} .
\end{equation}
Again, this is just a special case: the entire discussion can be generalized to phases which
are not necessarily plane waves.
In analogy with the case of gauge invariant (U(1)) field theories, we introduce a sort of gauge transformed version of the
flux operator:
\begin{equation}
\EE^i_{x_0} := E^{i} - v^{i}_{x_0}(h).
\end{equation}

These operators are just giving the group representation of the operators corresponding to the translated fluxes in the algebra
representation, with the residual dependence on the group element $h$ being 
associated to the noncommutative nature of such a translation, once the structure of
$\su$ is appropriately taken into account.

These new operators are obviously constructed in such a way that:
\begin{equation}
\EE^i_{x_0} U_{x_0}(f) = U_{x_0}( E^if),\qquad \EE^i_{x_0} = U_{x_0} E^i U_{x_0}^{\dagger}.
\end{equation}
Using this definition, one immediately realizes that the algebra of the $\EE^i$ is actually
the $\su$ one,  
\begin{equation}
[\EE^{i},\EE^j] = [U_{x_0} E^i U_{x_0}^{\dagger},U_{x_0} E^j U_{x_0}^{\dagger}]
= \iu \epsilon_{ijk} U_{x_0} E^{k} U_{x_0}^{\dagger} =
\iu \epsilon_{ijk} \EE^{k},
\end{equation}
as expected from the fact that, in flux space, the $\EE^i$ are just translated fluxes, still belonging to
$\su$.
This definition allows us to understand better what these new coherent states are. Define the
Casimir operator
\begin{equation}
\XX_{x_0} = -\delta_{ij}\EE^i_{x_0} \EE^j_{x_0}.
\label{laplacian}
\end{equation}
Due to the properties elucidated in the previous section, this operator, a generalized Laplacian,
is isospectral to the Laplace operator on $\SU$, with the functions $\DD^l$ providing an 
orthonormal diagonalizing basis.
Indeed:
\begin{eqnarray}
\XX_{x_0} \DD^{l}_{ab}
& = &
\delta_{ij}\EE_{x_0}^i \EE_{x_0}^j \DD^{l}_{ab}
\nonumber \\
& = &
\delta_{ij}\EE^i_{x_0} U_{x_0}(E^j D^{l}_{ab}) = U_{x_0}( \delta_{ij} E^i E^j D^l_{ab}) = U_{x_0}(\triangle_{\SU} D^{l}_{ab}) = -l(l+1) U_{x_0}(D^{l}_{ab})
\nonumber \\
& = &
-l(l+1)\DD^l_{ab}.
\end{eqnarray}

Of course this is an obvious result from the perspective of gauge field theories: we are using a Laplacian obtained by squaring a covariant derivative, in the case in which the connection
one-form is a pure gauge.
For this reason, we can immediately conclude that the flux coherent states are solutions of the 
heat equation
\begin{equation}
\partial_{t} \KK{t}{g_0,x_0}(h) = \XX_{x_0} \KK{t}{g_0,x_0}(h),
\end{equation}
with initial condition
\begin{equation}
\KK{t=0}{g_0,x_0}(h) = \delta(hh_0^{-1}) e_{h}(x_0) .
\end{equation}

This is the main result of the paper: the flux coherent states that were introduced in \cite{OPS1} are indeed heat kernel states,
with the difference that now the Laplacian entering the heat equation is not the Laplace--Beltrami
operator of $\SU$, but a more general operator that includes the effect of the translation
in the Lie algebra direction of the phase space. This is the group representation counterpart
of the statement contained in \eqref{eq:hkfourier}. Notice that the operator $\XX$ depends on $x_0$, and hence to obtain coherent states peaked on different Lie algebra elements we
have to use heat kernel associated to different Laplacians. This is a key difference with respect to
the analytically continued heat kernels, that are defined with respect to the same Laplace--Beltrami
operator, whatever is the position of their peak.


\section{Generalizations}
The results discussed so far can be made even more clear with a more suggestive notation. Let us introduce ket vectors, labeled by $\SU$ group elements,
\begin{equation}
\ket{h_0}{}: \qquad \langle h| h_0 \rangle = \delta(hh_0^{-1}),
\end{equation}
such that the heat kernel coherent states and the flux coherent states, denoted respectively as
\begin{equation}
\oket{h_0}{t} = e^{t\triangle}\ket{h_0}{}, \qquad \ket{h_0,x_0}{t} = e^{t\XX_{x_0}} U_{x_0} \ket{h_0}{},
\end{equation}
have wavefunctions
\begin{equation}
\HK{t}{h_0}(h) = \langle h \oket{g_0}{t} = \bra{h}{}\exp(t\triangle) \ket{h_0}{} ,
\qquad
\KK{t}{g_0,x_0}(h) = \langle h \ket{h_0,x_0}{t},
\end{equation}
Using the relation with the modified/translated flux operator,
\begin{equation}
\XX_{x_0} U_{x_0} = U_{x_0} \triangle,
\end{equation}
we obtain immediately:
\begin{equation}
\ket{h_0,x_0}{t} = e^{t\XX} U_{x_0} \ket{h_0}{} =U_{x_0} e^{t\triangle}  \ket{h_0}{} = U_{x_0} \oket{h_0}{t},
\end{equation}
which is exactly what we have already obtained, written in a different way. This rewriting allows us to
show immediately how to generalize the results obtained previously to different operators $U_{x_0}$, possibly not plane waves, and to other groups.
This rewriting also shows how the expectation values of operators can then be immediately inferred from the ones of the standard heat kernel,
once we remember that
$
h U_{x_0} = U_{x_0} h$ and $  \EE^{i}_{x_0}= U_{x_0} E^{i} U_{x_0}^{\dagger}
$:
\begin{eqnarray}
\langle \varphi(h) \rangle & = &  
\bra{h_0,x_0}{t} \varphi(h) \ket{h_0,x_0}{t} =
\obra{h_0}{t} \varphi(h) \oket{h_0}{t} \, ;
\\
\langle E^i \rangle &
= &
\bra{h_0,x_0}{t} E^i \ket{h_0,x_0}{t} 
=
\obra{h_0}{t} U^{\dagger} E^{i} U \oket{h_0}{t} 
=
\obra{h_0}{t} \EE^{i}_{-x_0}\oket{h_{0}}{t} = 0 + \langle v^{i}(x_0) \rangle_{h_0;t}\,;
\\
\langle E^{i}E^{j} \rangle &
= &
\bra{h_0,x_0}{t} E^i E^{j} \ket{h_0,x_0}{t} 
=
\obra{h_0}{t} U^{\dagger} E^{i} E^{j} U \oket{h_0}{t} 
=
\obra{h_0}{t} \EE^{i}_{-x_0}\EE^{j}_{-x_0}\oket{h_{0}}{t} \,. 
\end{eqnarray}
Hence, all the statistical properties, and in particular the behaviour with respect to the Heisenberg
uncertainty relations, are consistently imported from the ones determined for
the Hall states, as expected. 

In \cite{ThomasComplexifier} it was further considered the construction of the heat kernels in
term of complexifiers and the associated creation/annihilation operators. In particular, it has been
shown that it is natural to identify the annihilation operator with
\begin{equation}
A = e^{t\triangle} h e^{-t\triangle},
\end{equation}
whose eigenvectors are indeed the heat kernels. In analogy with that case, we can construct similar annihilation operators for our case by replacing the Laplacian $\triangle$ with
the modified one $\XX_{x_0}$. This is possible because the two operators possess the same spectrum; indeed, they are related by a unitary transformation.
In a similar way, we can associate to the Laplacian \eqref{laplacian} the annihilation operator
\begin{equation}
A(x_0) = e^{t\XX} h e^{-t\XX},
\end{equation}
that by construction treats the flux coherent states as its eigenvector.


\section{Concluding remarks}
Let us briefly summarize the previous points.
We have elucidated the structural similarities of the proposed flux coherent states with the complexifier coherent states, by their explicit representation in terms of group variables and
operators acting on $\lsq$.

We have explicitly shown the reason of their coherent behavior by tracing it back to their
role as heat kernels for a modified Laplace operator, $\XX$ isospectral to $\triangle_{\SU}$. In
particular, this might be used to reproduce the complexifier structure introduced for heat kernels.
Furthermore, the mapping between these states and the heat kernels ensures us that the statistical
properties of the latter will be imported automatically, allowing us to use them as coherent states.

At the same time, we have given a general recipe for the construction of coherent states based
on a generalized notion of heat kernel on the group manifold, by highlighting the nature of the operation that is at the basis
of the definition of the new states.
Indeed, given any family of unitary maps $U$, if we define
\begin{equation}
\EE^i = U E^{i} U^{\dagger}, \qquad \XX = -\delta_{ij}\EE^i\EE^j,
\end{equation}
we can construct arbitrary families of coherent states, provided that i) the map $U$ can be
parametrized by $\su$ elements, to which it will be associated the position of the peak of
the state in the Lie algebra, ii) that the map between the coherent states and $T^*\SU$
defined by the expectation values of $\hat{h}$ and $\hat{E}$ is bijective, and iii) provided that the overcompleteness and overlap properties
customarily valid fo coherent states are satisfied.

These indeed are the only conditions to be really checked explicitly as the conditions restricting
the maps $U$ to be of a specific form. 
For them one really
needs to understand the dependence of the operators $\exp(t\XX)$ on the Lie algebra elements.
While the proof of overcompleteness might be only a technical problem to find a measure with respect 
to which perform the integration in the Lie algebra elements (taken care of by the star product,
in the specific case of the flux coherent states), the proof of overlap properties requires the understanding of integrals
of the form
\begin{equation}
_t\langle g_0, y_0 \ket{h_0,x_0}{t}= \obra{g_0}{t} U^{\dagger}_{y_0}U_{x_0} \oket{h_0}{t} =
\obra{g_0}{t} U_{x_0-y_0} \oket{h_0}{t},
\end{equation}
for which of course the explicit form of the operators $U$ is needed. \textit{A posteriori},
in \cite{OPS1} these were the only two properties that needed to be verified,
the others following in a straightforward manner from the properties of the heat kernel,
already established in the literature.

\vspace{0.3cm}

\textbf{Acknowledgments}: The authors would like to thank S. Ansoldi for remarks on an earlier
version of the draft and D. Oriti for fruitful discussions.


\bibliographystyle{unsrt}
\bibliography{biblio}

\end{document}